\documentclass[a4paper,fleqn,usenatbib]{mnras}

\usepackage[T1]{fontenc}
\usepackage{ae,aecompl}

\usepackage{graphicx}	
\usepackage{amsmath}	
\usepackage{amssymb}	
\usepackage{mathrsfs}                

\title[Strongly Lensed GWs and EM Signals as Cosmic Rulers]{Strongly Lensed Gravitational Waves and Electromagnetic Signals as Powerful Cosmic Rulers}

\author[Wei \& Wu]{
Jun-Jie Wei$^{1,2}$\thanks{E-mail: jjwei@pmo.ac.cn (JJW)}
and Xue-Feng Wu$^{1,3,4}$\thanks{E-mail: xfwu@pmo.ac.cn (XFW)}
\\
$^{1}$Purple Mountain Observatory, Chinese Academy of Sciences, Nanjing 210008, China\\
$^{2}$Guangxi Key Laboratory for Relativistic Astrophysics, Nanning 530004, China\\
$^{3}$School of Astronomy and Space Sciences, University of Science and Technology of China, Hefei 230026, China\\
$^{4}$Joint Center for Particle, Nuclear Physics and Cosmology, Nanjing University-Purple Mountain Observatory, Nanjing 210008, China
}

\date{Accepted XXX. Received YYY; in original form ZZZ}

\pubyear{2015}

\begin{document}
\label{firstpage}
\pagerange{\pageref{firstpage}--\pageref{lastpage}}
\maketitle

\begin{abstract}
In this paper, we discuss the possibility of using strongly lensed gravitational waves (GWs) and their electromagnetic
(EM) counterparts as powerful cosmic rulers. In the EM domain, it has been suggested that joint observations of
the time delay ($\Delta\tau$) between lensed quasar images and the velocity dispersion ($\sigma$) of the lensing galaxy
(i.e., the combination $\Delta\tau/\sigma^{2}$) are able to constrain the cosmological parameters more strongly than
$\Delta\tau$ or $\sigma^{2}$ separately. Here, for the first time, we propose that this $\Delta\tau/\sigma^{2}$ method can be applied to
the strongly lensed systems observed in both GW and EM windows. Combining the redshifts, images and $\sigma$ observed
in the EM domain with the very precise $\Delta\tau$ derived from lensed GW signals, we expect that accurate multimessenger
cosmology can be achieved in the era of third-generation GW detectors. Comparing with the constraints from
the $\Delta\tau$ method, we prove that using $\Delta\tau/\sigma^{2}$ can improve the discrimination between cosmological models.
Furthermore, we demonstrate that with $\sim50$ strongly lensed GW-EM systems, we can reach a constraint on the dark energy
equation of state $w$ comparable to the 580 Union2.1 Type Ia supernovae data. Much more stringent constraints on $w$ can be
obtained when combining the $\Delta\tau$ and $\Delta\tau/\sigma^{2}$ methods.
\end{abstract}

\begin{keywords}
gravitational lensing: strong -- gravitational waves -- cosmological parameters
-- dark energy -- distance scale
\end{keywords}



\section{Introduction}
\label{sec:Intro}

In the standard $\Lambda$CDM cosmological model, it is currently inferred that $\sim96\%$ of the total energy density
of the Universe consists of dark matter ($\sim26\%$) and dark energy ($\sim70\%$). These proportions have
been measured precisely via various standard candles or rulers, such as Type Ia supernovae (SNe Ia;
\citealt{1998Natur.391...51P,1998AJ....116.1009R,1998ApJ...507...46S,2012ApJ...746...85S}), anisotropy of the
cosmic microwave background (CMB) radiation \citep{2013ApJS..208...19H,2016A&A...594A..13P}, as well as
baryon acoustic oscillations \citep{2011MNRAS.416.3017B,2012MNRAS.427.3435A}. Though cosmology has entered
a new era of precision tests, we should note that all of the cosmological probes are based on
electromagnetic (EM) observations alone.
In 1986, \citet{1986Natur.323..310S} first proposed that the waveform signal of gravitational waves (GWs)
from inspiralling and merging compact binaries encodes the luminosity distance $d_{L}$ information, proving
access to the direct measurement of $d_{L}$. Thus, the GW signals can be considered as standard sirens.
The combination of $d_{L}$ derived from GWs and redshifts $z$ derived from their EM counterparts would
make GW events an ideal tool to constrain the cosmological parameters and the equation of state of dark energy.

On 2016 February 11, the Laser Interferometer Gravitational Wave Observatory (LIGO) team reported the first direct
detection of the gravitational wave source (GW 150914; \citealt{2016PhRvL.116f1102A}), opening a brand
new window for studying the Universe, which indicates that the era of multimessenger cosmology is coming.
In the past, several studies have investigated the possibility of GWs as standard sirens
(e.g. \citealt{2005ApJ...629...15H,2017PhRvD..95d4024C,2017PhRvD..95d3502D}).
Particularly, \citet{2017PhRvD..95d4024C} found that with about 500--600 simulated GW events they can
determine the Hubble constant $H_{0}$ with an accuracy comparable to \emph{Planck} 2015 results; for the dark
matter density parameter $\Omega_{\rm m}$, it should need more than 1000 GW events to match the \emph{Planck} sensitivity.

Very recently, \citet{2017PhRvL.118i1102F} presented a new model-independent method for constraining the speed of GWs,
based on future time delay measurements of strongly lensed GWs and their EM counterparts (see also \citealt{2017PhRvD..95f3512B,2017PhRvL.118i1101C}).
Even more encouragingly, \citet{2017arXiv170304151L} have shown that such strongly lensed GW-EM systems could also provide strong constraints
on cosmological parameters. The GW standard-sirens method in cosmology appeals to the luminosity distance measurement
from the GW observation which relies on the fine details of the waveform, but the proposed method of \citet{2017arXiv170304151L}
is waveform independent. Moreover, the GW standard-sirens method would require several hundred GW events to match the
\emph{Planck} sensitivity on $H_{0}$ \citep{2017PhRvD..95d4024C}, while \citet{2017arXiv170304151L}
have shown that the uncertainty of $H_{0}$ might be better constrained by future time delays of lensed GW-EM events.
Note that the time delays between different lensed images ($\sim10-100$ days) obtained from the GW observations
would reach an unprecedented accuracy of $\sim0.1$ s from the detection pipeline. Compared to the uncertainty in lens modelling,
the uncertainty of the GW time delay is negligible.

Strong gravitational lenses are a complementary cosmological probe (see e.g.
\citealt{1964MNRAS.128..307R,2000IJMPD...9..591Z,2008A&A...477..397G,2010MNRAS.406.1055B,2010ARA&A..48...87T,
2012A&A...538A..43C,2012JCAP...03..016C,2012ApJ...755...31C,2015ApJ...806..185C,2012AJ....143..120O,2014MNRAS.443..969C}).
The Einstein ring radius inferred from the deflection angle and the time delay between different lensed images can provide the information
of angular-size distance independently, which can be used to measure cosmological parameters (see e.g.
\citealt{2009ApJ...706...45C,2009MNRAS.397..311D,2009A&A...507L..49P,2010ApJ...712.1378P,2010ApJ...711..201S,2013ApJ...766...70S,
2014MNRAS.437..600S,2017MNRAS.465.4914B}), to discriminate different cosmological models
(see, e.g., \citealt{2008A&A...487..831Z,2014ApJ...788..190W,2015AJ....149....2M,2015MNRAS.452.2423Y}),
and to probe the cosmic distance duality relation (see, e.g., \citealt{2016JCAP...02..054H,2016ApJ...822...74L,2017JCAP...07..010R}).
As of today, the observation of about 70
strong gravitational lensing systems and 12 two-image lensing systems with time delay measurements have provided the data that,
in principle, can be used to carry out the study of cosmology. However, this is only the beginning. The upcoming Large Synoptic Survey
Telescope (LSST) will find more than $\sim8000$ lensed quasars, about $3000$ of which will have well-measured time delays within 10 yr \citep{2010MNRAS.405.2579O}.
The number of robust time-delay measurements for probing cosmology is estimated to be $\sim400$, each with precision $<3\%$ and accuracy of
$\sim1\%$ \citep{2015ApJ...799..168D,2015ApJ...800...11L}. In addition, we note that \cite{2009A&A...507L..49P} have proposed
an interesting cosmic ruler constructed from the joint measurements of the time delay ($\Delta\tau$) between lensed quasar images
and the velocity dispersion ($\sigma$) of the lensing galaxy. They have shown that the joint measurement of $\Delta\tau/\sigma^{2}$
is more effective to constrain cosmological parameters than $\Delta\tau$ or $\sigma^{2}$ separately.

In the GW window, the fantastic sensitivity of the third-generation GW interferometric detectors, such as the \emph{Einstein} Telescope (ET),
would significantly improve the detection efficiencies of the GW events. With a large number of detectable events, we might expect
some of these events to be gravitationally lensed by intervening galaxies. The prospects of observing strongly lensed GWs from
merging double compact objects (NS-NS, NS-BH, BH-BH) have been studied in detail \citep{2013JCAP...10..022P,2014JCAP...10..080B,2015JCAP...12..006D};
these works have predicted that ET would detect about 50--100 strongly lensed GW events per year. This implies that the ET will be able to
provide a considerable catalogue of strongly lensed GWs within a few years of successful operation.

As mentioned above, \cite{2017arXiv170304151L} proposed that future time delay measurements ($\Delta\tau$) of strongly lensed GW signals accompanied
by EM counterparts could be used to obtain robust constraints on cosmological parameters. Because $\Delta\tau/\sigma^{2}$ is more sensitive to
the cosmological parameters than $\Delta\tau$ or $\sigma^{2}$ separately \citep{2009A&A...507L..49P}, here, for the first time,
we try to explore the cosmological constraint ability by future joint measurements of the precise time delay ($\Delta\tau$) between lensed GW images and the velocity dispersion
($\sigma$) of the lensing galaxy in the era of the third-generation GW detectors.

The paper is organized as follows. In Section~\ref{sec:lenses}, we describe the basics of using strong gravitational lensing systems
as standard rulers. In Section~\ref{sec:MC}, we demonstrate that the cosmological parameters can be constrained with great accuracy
through the combination $\Delta\tau/\sigma^{2}$ of the lensed GW-EM system, using Monte Carlo simulations.
A brief summary and discussion are given in Section~\ref{sec:summary}.

\section{Strong lenses as cosmic rulers}
\label{sec:lenses}

A source lensed by a foreground massive galaxy or galaxy cluster appears in multiple images.
For a given image $i$ at angle position $\vec{\theta}_i$, with the source position at angle $\vec\beta$,
the time delay $\Delta\tau_i$ is caused both by the difference in path-length between the straight and deflected rays,
and the gravitational time dilation of the light ray traveling through the effective gravitational potential
$\Psi(\vec\theta_i)$ of the lens \citep{1986ApJ...310..568B}:
\begin{equation}
\Delta\tau_i=\frac{1+z_{\rm L}}{c}\frac{D_{\rm OS}D_{\rm OL}}{D_{\rm LS}}\left[\frac{1}{2}(\vec{\theta_i}-\vec{\beta})^2-\Psi(\vec{\theta_i})\right]\;.
\end{equation}
Here, $z_{\rm L}$ is the lens redshift and $D_{\rm OS}$, $D_{\rm OL}$, and $D_{\rm LS}$ represent the angular-diameter distances
between observer and source, observer and lens, and lens and source, respectively. If the lens potential $\Psi$
and the lens geometry $\vec{\theta_i}-\vec{\beta}$ are known, the time delay measures the ratio $D_{\rm OS}D_{\rm OL}/D_{\rm LS}$,
which depends on the cosmological parameters.
Assuming that the time-delay lensing systems have only two images at $\vec\theta_A$ and $\vec\theta_B$,
and adopting the single isothermal sphere (SIS) model for the gravitational potential of the lens galaxy,
the time delay is therefore given by
\begin{equation}
\Delta\tau=\frac{1+z_{\rm L}}{2c}\frac{D_{\rm OS}D_{\rm OL}}{D_{\rm LS}}(\theta_B^2-\theta_A^2)\;.
\label{eq:time delay}
\end{equation}
The distance ratio that appears in Equation~(\ref{eq:time delay}) is the time-delay distance,
$D_{\Delta \tau}\equiv(1+z_{\rm L})D_{\rm OS}D_{\rm OL}/D_{\rm LS}$, which depends primarily on $H_{0}$
and has a limited sensitivity to other cosmological parameters, such as $\Omega_{\rm m}$ (more on this below).

\begin{figure*}
\begin{center}
\includegraphics[width=1.0\textwidth]{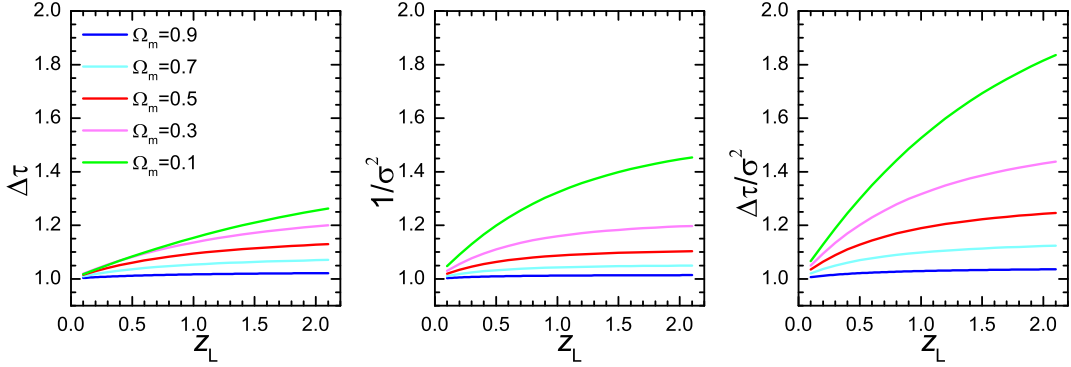}
\caption{Sensitivity of the three methods ($\Delta\tau$, $\sigma^2$, and $\Delta\tau/\sigma^2$) to the cosmological parameters.
The source redshift $z_{\rm S}$ is fixed to 3. A flat Universe is assumed with five different $\Omega_{\rm m}$ values: $0.1$,
$0.3$, $0.5$, $0.7$, and $0.9$. Each curve is calculated relative to the Einstein-de Sitter Universe.}
\vskip-0.2in
\label{fig:f1}
\end{center}
\end{figure*}

Inferring cosmological distances from time-delay lenses also requires accurate models for the mass distribution of the lens galaxy,
as well as for any other matter structures along the line-of-sight that might affect the
observed time delays between the multiple images \citep{2010ApJ...711..201S}. A constant external convergence term
$\kappa_{\rm ext}$ can be absorbed by the lens and source model, leaving the fit to the lensed images unchanged.
However, the true time-delay distance $D_{\Delta \tau}$ is altered by a factor of $(1-\kappa_{\rm ext})$, i.e.,
\begin{equation}
D_{\Delta \tau}=\frac{D^{(0)}_{\Delta \tau}}{1-\kappa_{\rm ext}}\;,
\label{eq:line-of-sight}
\end{equation}
where $D^{(0)}_{\Delta \tau}$ is the time-delay distance inferred from a model not accounting for the effects of
weak perturbers along the line-of-sight. To break the ``mass-sheet degeneracy'' \citep{1985ApJ...289L...1F},
it is possible to study the lens environment to constrain $\kappa_{\rm ext}$ within a few percent based on spectroscopy and
multiband wide-field observations of local galaxy groups and line-of-sight structures (e.g., \citealt{2006ApJ...642...30F,2006ApJ...641..169M})
in combination with ray-tracing through numerical simulations (e.g., \citealt{2013MNRAS.432..679C,2013ApJ...768...39G}).
According to the recent analysis by \citet{2016MNRAS.462.3255C}, the external convergence over an ensemble of lenses usually
does not average to zero. \citet{2017MNRAS.467.4220R} presented a robust estimate of the external convergence $\kappa_{\rm ext}$
for the lensed quasar HE 0435-1223, which has a median of 0.004 and a standard deviation of $\delta_{\kappa_{\rm ext}}=0.025$.
This measured $\delta_{\kappa_{\rm ext}}$ corresponds to 2.5\% uncertainty on $D_{\Delta \tau}$. In sum, the external convergence
of each lens is expected to introduce 1 or 2 percent extra uncertainty on $D_{\Delta \tau}$
\citep{2013MNRAS.432..679C,2013ApJ...768...39G,2017MNRAS.467.4220R}. Thus, the uncertainty on $D_{\Delta \tau}$ is
given by the quadrature sum of the uncertainties on the time delay, external convergence, and image position measurements
\citep{2017MNRAS.468.2590S}.

The observed velocity dispersion ($\sigma$) of the lensing galaxy is the result of the superposition of
numerous individual stellar spectra, each of which has been Doppler shifted because of the random stellar
motions within the galaxy. Hence, it can be measured by analysing the integrated spectrum of the galaxy.
According to the virial theorem, the velocity dispersion is related to the mass (i.e. $\sigma^{2}\propto M_{\sigma}R$,
where $M_{\sigma}$ denote the mass enclosed inside the radius $R$). The mass is determined by the Einstein
ring radius $\theta_{\rm E}$ of the lensing system, and thus the velocity dispersion in the SIS model can be
written as
\begin{equation}
\sigma^2=\theta_{\rm E}\frac{c^2}{4\pi}\frac{D_{\rm OS}}{D_{\rm LS}}\;.
\label{eq:velocity}
\end{equation}

As shown by \cite{2009A&A...507L..49P}, two of the angular-diameter distances appearing in Equation~(\ref{eq:time delay})
could be replaced by the velocity dispersion $\sigma$ and the Einstein radius $\theta_{\rm E}=(\theta_A+\theta_B)/2$
($\Delta\tau$ and $\theta_{A,B}$ are defined to be positive here, with $\theta_B>\theta_A$), i.e.
\begin{equation}
D_{\rm OL}(\theta_B-\theta_A)=\frac{c^3}{4\pi}\frac{\Delta\tau}{\sigma^2(1+z_{\rm L})}\;.
\label{eq:combine}
\end{equation}
In contrast to $D_{\Delta \tau}$, \cite{2015JCAP...11..033J} found that the mass external to the lens along the
line-of-sight (external convergence) has no effect on the inferred $D_{\rm OL}$. The reason is as follows.
Assume that there is a lens system which has a time delay of $\Delta\tau$ and a velocity dispersion of $\sigma^2$.
We then try to model this system by a lens plus an external convergence, $\kappa_{\rm ext}$. The modelled $\Delta\tau$
and $\sigma^2$ would be different from the original ones by a factor of $(1-\kappa_{\rm ext})$, but the ratio of the two
is invariant. Because $D_{\rm OL}$ is proportional to the ratio $\Delta\tau/\sigma^2$, we can determine the same $D_{\rm OL}$
as before, regardless of the existence of the external convergence. Thus, the uncertainty on $D_{\rm OL}$ is
given by the quadrature sum of the uncertainties on the time delay, velocity dispersion, and image position measurements
\citep{2015JCAP...11..033J,2016JCAP...04..031J}.

From Equations~(\ref{eq:time delay}), (\ref{eq:velocity}), and (\ref{eq:combine}), we can see that
the time-delay $\Delta\tau$ is proportional to $D_{\rm OS}D_{\rm OL}/D_{\rm LS}$, the square of the velocity dispersion $\sigma^2$
is proportional to $D_{\rm OS}/D_{\rm LS}$, and the ratio $\Delta\tau/\sigma^2$ is dependent only on $D_{\rm OL}$. That is to say,
\begin{equation}
\Delta\tau\propto\frac{D_{\rm OS}D_{\rm OL}}{D_{\rm LS}},\;\sigma^2\propto\frac{D_{\rm OS}}{D_{\rm LS}},\;\frac{\Delta\tau}{\sigma^2}\propto {D_{\rm OL}}\;.
\label{eq:rulers}
\end{equation}
In Fig.~\ref{fig:f1}, we show the three quantities ($\Delta\tau$, $\sigma^2$, and $\Delta\tau/\sigma^2$) as a function of the lens redshift $z_{\rm L}$
in the flat $\Lambda$CDM model with a fixed source redshift $z_{\rm S}=3$ (see also \citealt{2009A&A...507L..49P}).
To illustrate the sensitivity of the three functions to $\Omega_{\rm m}$, we plot them for several cases of a flat Universe with
$\Omega_{\rm m}=0.1,\;0.3,\;0.5,\;0.7$ and $0.9$, relative to the Einstein-de Sitter Universe ($\Omega_{\rm m}=1,\;\Omega_{\Lambda}=0$).
We can see from this plot that the $\Delta\tau/\sigma^2$ curves have a wider separation than the $\Delta\tau$ or $\sigma^2$ curves
to allow an easier discrimination among different cosmological models. This is especially so at high redshifts, and thus it is
of special significance for the $\Delta\tau/\sigma^2$ method to study high-redshift lenses. Moreover, this method has the
advantage of being independent of the source redshift.
However, before we put the $\Delta\tau/\sigma^2$ method into practical cosmological use, we must consider its independence from
the other two methods. From their independence tests, \cite{2015MNRAS.452.2423Y} found that the methods of $\Delta\tau/\sigma^2$
and $\sigma^2$ are not independent, but $\Delta\tau/\sigma^2$ and $\Delta\tau$ are independent; thus, we can compare the capability
of these two methods to conduct cosmography.

\section{Testing the capability of lensed GW-EM events to conduct cosmography}
\label{sec:MC}

\subsection{Monte Carlo simulations}

We perform Monte Carlo simulations to test how well the two quantities ($\Delta\tau$ and $\Delta\tau/\sigma^{2}$)
from strongly lensed GW-EM systems can be used to constrain cosmological parameters.
To do so, we have to choose a fiducial cosmological model and then simulate a sample of lensed GW-EM systems.
Here we adopt the following cosmological parameters of the flat $\Lambda$CDM model derived from \emph{Planck} 2015 data \citep{2016A&A...594A..13P}
in our simulations: $H_{0}=67.8$ km $\rm s^{-1}$ $\rm Mpc^{-1}$, $\Omega_{\rm m}=0.308$, and $\Omega_{\Lambda}=1-\Omega_{\rm m}$.
Our detailed simulation steps are described as follows:

1. The redshifts of source $z_{\rm S}$ and lens $z_{\rm L}$ are randomly generated from the expected redshift probability distribution functions (PDFs)
of lensed GW events \citep{2014JCAP...10..080B,2015JCAP...12..006D}. These redshift PDFs were calculated using the following procedure.
First, considering the intrinsic merger rates of the whole class of double compact objects located at different reshifts
as calculated by \cite{2013ApJ...779...72D} and the designed sensitivity of the ET, the yearly detection rate of GW events was estimated. Secondly, the probability
that individual GW signals from inspiralling double compact objects could be lensed by an early-type galaxy was then calculated.
Finally, adding all the double-compact-objects merging systems together, the yearly detection rate of lensed GW events
detected by the ET was predicted. This prediction is accompanied by the redshift PDF (see Fig.~2 of \cite{2015JCAP...12..006D}),
which enables us to randomly generate the samples of $z_{\rm S}$ and $z_{\rm L}$.
Note that short gamma-ray bursts (short GRBs), on-beam GRB afterglow emission, and kilonovae/mergenovae are considered as promising EM
counterparts of GW signals. Because $z<3$ for the current short GRBs, the range of the source redshift $z_{\rm S}$ for our analysis
is from 0 to 3.

2. We simulate the velocity dispersion $\sigma$ and time delay $\Delta\tau$ separately from the probability distributions
of $\sigma$ and $\Delta\tau$ from the OM10 catalogue \citep{2010MNRAS.405.2579O}. The OM10 catalogue provides mock observations of lensed
quasars expected for the baseline survey planned with the LSST, based on realistic distributions of quasars and elliptical galaxies as well as
the observational condition of this telescope.

3. For the $\Delta\tau$ method, the mock external convergence $\kappa_{\rm ext}$ is obtained by sampling the PDF of $\kappa_{\rm ext}$
given by \cite{2017MNRAS.467.4220R}. We then infer the fiducial value of $\Theta\equiv(\theta_B^2-\theta_A^2)$ from Equations~(\ref{eq:time delay}) and (\ref{eq:line-of-sight})
using the mock $z_{\rm S}$, $z_{\rm L}$, $\kappa_{\rm ext}$, and $\Delta\tau$. For the $\Delta\tau/\sigma^{2}$ method, the fiducial value of $\Delta\theta\equiv(\theta_B-\theta_A)$
is inferred from Equation~(\ref{eq:combine}) with the mock $z_{\rm L}$, $\sigma$, and $\Delta\tau$.

4. Large numbers of new strong gravitational lenses will be discovered by dedicated surveys,
including the LSST project \citep{2011AAS...21725222M,2013MNRAS.434.2121C}, the Dark Energy Survey
(DES; \citealt{2008MNRAS.386.1219B,2014AAS...22324801B,2014PhRvL.112f1301S}), and the VST ATLAS survey \citep{2014MNRAS.442L..85K}.
Also, time delays will be accurately constrained for a subsample of these with subsequent monitoring observations.
The precision of time-delay measurements is estimated to be $<3\%$ \citep{2015ApJ...799..168D,2015ApJ...800...11L}.
While the dedicated observations of lensed quasar systems in the EM domain give $\sim3\%$ uncertainty of
the time-delay measurement \citep{2015ApJ...800...11L}, $\Delta\tau$ obtained from the GW signals are supposed
to be very accurate with negligible uncertainty. Therefore, we assign an uncertainty $\delta_{\Delta\tau}=3\%\Delta\tau$
to the lensed quasar system, and $\delta_{\Delta\tau}\simeq0$ to the lensed GW-EM system.

5. The current techniques concerning lensed quasar systems in the EM window give a few percent uncertainty of the determination of
lens modelling, that is, $\sim4\%$ uncertainty on image position measurement $\Theta$ (or $\Delta\theta$) and $\sim10\%$ uncertainty
for the observed velocity dispersion $\sigma^{2}$ \citep{2015JCAP...11..033J}.
Note that the bright point spread functions (PSFs) of active galaxy nuclei (AGNs) could induce large systematic errors.
In order to extract bright AGN images during the lens modelling procedure, instead one has to use a nearby star's PSF or to adopt an iterative PSF
modelling process, which can accurately recover the PSF for real observations \citep{2016MNRAS.462.3457C,2017MNRAS.465.4634D,2017MNRAS.465.4895W}.
However, the systematic errors can not be completely eliminated by these operations. Unlike AGNs, the EM counterparts of GW signals,
such as short GRBs and kilonovae, are not always so bright. Therefore, the lensed images might not be affected much by the bright PSFs,
which are difficult to extract, and the exposure time could be longer, making lens modelling so much easier.
Based on the current lensing project H0LiCOW\footnote{http://h0licow.org}, \cite{2017arXiv170304151L} simulated two sets of
realistic lensed images with and without the AGN (see \citealt{2017MNRAS.465.4634D} for more details on the simulations).
The corresponding exposure time and noise level were set as close as possible the deep Hubble Space Telescope observations.
They found that the effect of bright PSFs can significantly influence the uncertainties of the parameters in the lens model.
Therefore, \cite{2017arXiv170304151L} suggested that the accuracy of lens modelling would be improved to some extent with gravitationally lensed GWs and EM signals.
For each lensed GW-EM event, we assign the uncertainties $\delta_{\Theta}=2\%\Theta$
(or $\delta_{\Delta\theta}=2\%\Delta\theta$) and $\delta_{\sigma^{2}}=5\%\sigma^{2}$ to the mock $\Theta$ (or $\Delta\theta$) and velocity dispersion $\sigma^{2}$
separately. These are supposed to be the best-case scenarios in the GW era, and they are two times smaller than those of the lensed quasar system in the EM domain.
With large-aperture telescopes (e.g. the Thirty Meter Telescope and the European Extremely Large Telescope) or the James Webb Space Telescope,
the required precision of velocity dispersions could be achieved. This should be coupled with high-resolution imaging, which can effectively constrain
the density structures of the lenses and mass structures affecting the lensing. Density structures and extrinsic mass can also be constrained by
modelling the lensing configuration, including positions and flux ratios of the images \citep{2009A&A...507L..49P}.

6. It should be underlined that lensed GWs might provide some help with improving lens modelling uncertainty, but they do not help with the uncertainty of external convergence $\kappa_{\rm ext}$.
This is because in order to accurately quantify the mass distribution along the line-of-sight, wide-field imaging and spectroscopy are required (see \citealt{2016A&ARv..24...11T} for a recent review),
either in lensed quasars or in lensed GWs. \cite{2017MNRAS.467.4220R} have shown that the uncertainty of $\kappa_{\rm ext}$ would
contribute a root-mean-square error of 1 or 2 percent to the value of $D_{\Delta \tau}$ (see also \citealt{2013MNRAS.432..679C,2013ApJ...768...39G}).
Thus, we assign an uncertainty $\delta_{\kappa_{\rm ext}}=1\%(1-\kappa_{\rm ext})$ to the mock $\kappa_{\rm ext}$ for both the lensed quasar system and the lensed GW-EM system.

7. For every synthetic lens, we add a deviation to the fiducial value of $\Theta^{\rm fid}$ (or $\Delta\theta^{\rm fid}$).
That is, we sample the $\Theta^{\rm mea}$ (or $\Delta\theta^{\rm mea}$) measurement according to the Gaussian distribution
$\Theta^{\rm mea}=\mathcal{N}(\Theta^{\rm fid},\;\sigma_{\Theta})$ (or $\Delta\theta^{\rm mea}=\mathcal{N}(\Delta\theta^{\rm fid},\;\sigma_{\Delta\theta})$).

8. Repeat the above steps to obtain a sample of 50 strong lenses.

\subsection{Estimation of cosmological parameters}

For a set of 50 simulated lenses, the likelihood for the cosmological parameters can be determined from the minimum $\chi^{2}$ statistic:
\begin{equation}
\chi^2(\mathbf{p})=\sum_{i}\frac{\left[\mathcal{D}_i^{\rm obs}-\mathcal{D}_{i}^{\rm th}(\mathbf{p})\right]^2}{\delta^2_{\mathcal{D}_{i}}}\;.
\end{equation}
Here, $\mathcal{D}^{\rm th}$ is the theoretical distance calculated from the set of cosmological parameters $\mathbf{p}$,
$\mathcal{D}^{\rm th}=D_{\rm OS}D_{\rm OL}/D_{\rm LS}$ and $\mathcal{D}^{\rm th}=D_{\rm OL}$ correspond to the $\Delta\tau$ method
and the $\Delta\tau/\sigma^{2}$ method, respectively. $\mathcal{D}^{\rm obs}$ is the distance of the simulated observational data sets
and $\delta_{\mathcal{D}}$ is the error of $\mathcal{D}^{\rm obs}$. With the measured distance $\mathcal{D}^{\rm obs}$
(see Equations~\ref{eq:time delay} and \ref{eq:line-of-sight}), the propagated error $\delta_{\mathcal{D}_{\Delta\tau}}$ in $\mathcal{D}^{\rm obs}$ using the $\Delta\tau$ method is
\begin{equation}
\delta_{\mathcal{D}_{\Delta\tau}}=\mathcal{D}^{\rm obs}\left[\left(\frac{\delta_{\Delta\tau}}{\Delta\tau}\right)^2+\left(\frac{\delta_{\Theta}}{\Theta}\right)^2
+\left(\frac{\delta_{\kappa_{\rm ext}}}{1-\kappa_{\rm ext}}\right)^{2}\right]^{1/2}\;.
\label{eq:sigma_Time}
\end{equation}
From Equation~(\ref{eq:combine}), the propagated error $\delta_{\mathcal{D}_{\Delta\tau/\sigma^{2}}}$ in $\mathcal{D}^{\rm obs}$ using the $\Delta\tau/\sigma^{2}$ method
can be written as
\begin{equation}
\delta_{\mathcal{D}_{\Delta\tau/\sigma^{2}}}=\mathcal{D}^{\rm obs}\left[\left(\frac{\delta_{\Delta\tau}}{\Delta\tau}\right)^2+\left(\frac{\delta_{\Delta\theta}}{\Delta\theta}\right)^2
+\left(\frac{\delta_{\sigma^{2}}}{\sigma^{2}}\right)^2\right]^{1/2}\;,
\label{eq:sigma_combine}
\end{equation}
which is dominated by the uncertainty of the velocity dispersion $\sigma^{2}$ \citep{2015JCAP...11..033J}.
To ensure the final constraint results are unbiased, we repeat this process 1000 times for each data set by using different noise seeds.

\begin{figure}
\begin{center}
\vskip-0.4in
\includegraphics[width=0.3\textwidth]{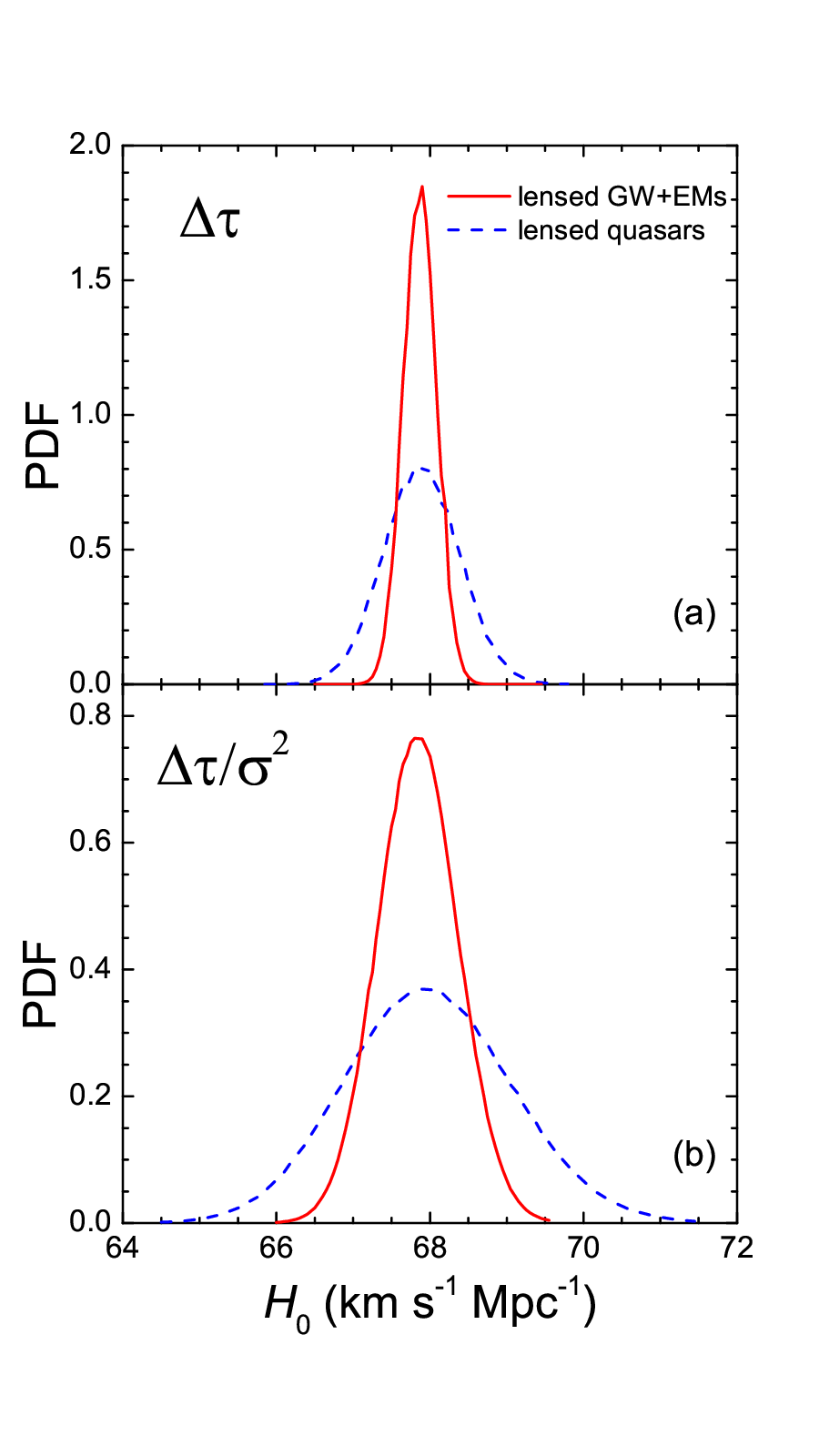}
\vskip-0.4in
\caption{Constraints on the Hubble constant, $H_{0}$, using 50 lensed GW-EM systems (red solid lines) and 50 lensed quasars (blue dashed lines):
(a) simulations for the $\Delta\tau$ method; (b) simulations for the $\Delta\tau/\sigma^{2}$ method.}
\vskip-0.2in
\label{fig:f2}
\end{center}
\end{figure}

In $\Lambda$CDM, the dark-energy equation-of-state parameter, $w$, is exactly $-1$.
Assuming a flat Universe, $\Omega_{\Lambda}=1-\Omega_{\rm m}$, there are only two free parameters: $\Omega_{\rm m}$ and $H_{0}$.
We first fix the flat $\Lambda$CDM model with $\Omega_{\rm m}=0.308$, but keep $H_{0}$ as a free parameter.
Fig.~\ref{fig:f2} shows the constraints on $H_{0}$ using two different quantities ($\Delta\tau$ and $\Delta\tau/\sigma^{2}$)
from 50 strongly lensed GW-EM systems (solid lines). For comparison, we also plot those constraints obtained from 50 lensed quasars (dashed lines) in the EM domain.
We can see that lensed systems observed jointly in GW and EM windows place much more stringent constraints on $H_{0}$ than pure EM lensed systems,
independent of what kind of observed quantity ($\Delta\tau$ or $\Delta\tau/\sigma^{2}$) is adopted.
This is mainly because the uncertainties of both the time delay and lens modelling in the lensed GW-EM systems
are smaller than those of the lensed quasar systems in the EM domain. Using $\Delta\tau$, we find that the uncertainty of $H_{0}$ from 50 lensed GW-EM systems
is $\sim0.3\%$, compared to $\sim0.7\%$ from pure EM lensed systems. Similarly, $H_{0}$ is better constrained by 50 lensed GW-EM systems
than by pure EM lensed systems with uncertainties of $\sim0.8\%$ versus $\sim1.6\%$ using $\Delta\tau/\sigma^{2}$.
Our results are in good agreement with \cite{2017arXiv170304151L}.
Not surprisingly, a comparison of Figs.~\ref{fig:f2}(a) and (b) shows that the $\Delta\tau/\sigma^{2}$ method gives weaker constraints
on $H_{0}$ than the other method, because the joint observations of time delay and velocity dispersion bring the extra uncertainty
from the velocity dispersion (see the comparison between Equations~(\ref{eq:sigma_Time}) and (\ref{eq:sigma_combine})).

\begin{figure}
\begin{center}
\vskip-0.4in
\includegraphics[width=0.3\textwidth]{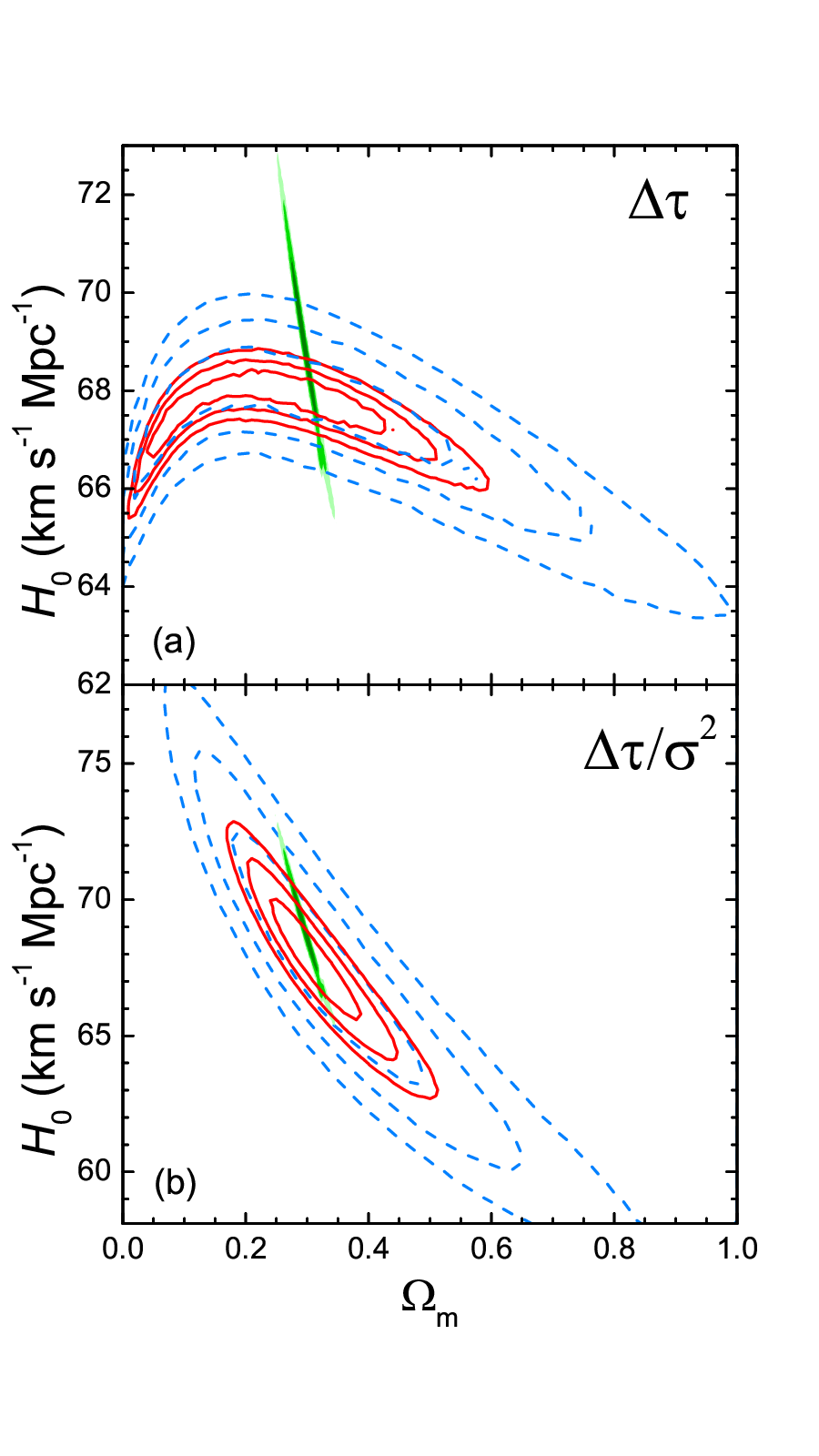}
\vskip-0.4in
\caption{The $1\sigma-3\sigma$ constraint contours of ($\Omega_{\rm m}$, $H_{0}$) in the flat $\Lambda$CDM model from
50 lensed GW-EM systems (red solid lines), 50 lensed quasars (blue dashed lines), and CMB data (green contours):
(a) simulations for the $\Delta\tau$ method; (b) simulations for the $\Delta\tau/\sigma^{2}$ method.}
\vskip-0.2in
\label{fig:f3}
\end{center}
\end{figure}

If we relax the priors, and allow both $H_{0}$ and $\Omega_{\rm m}$ to be free parameters, we obtain the constraints set
in the $\Omega_{\rm m}-H_{0}$ plane, as illustrated in Fig.~\ref{fig:f3}.
In the traditional approach using lensed quasar systems observed in the EM domain, we need a larger sample to increase the significance of the constraints.
In contrary, future observations of lensed GWs and their EM counterparts will enable us to achieve precise cosmography from around 50 such systems.
The constraints on the parameter space from the $\Delta\tau$ method (Fig.~\ref{fig:f3}(a)) give a good constraint on $H_{0}$,
but a weak constraint on $\Omega_{\rm m}$. As expected, the $\Delta\tau/\sigma^{2}$ method (Fig.~\ref{fig:f3}(b)) gives tighter constraints
on $\Omega_{\rm m}$ than the other method (i.e. the $\Delta\tau/\sigma^{2}$ method can improve the discrimination between cosmological models).

\begin{figure}
\begin{center}
\vskip-0.2in
\includegraphics[width=0.3\textwidth]{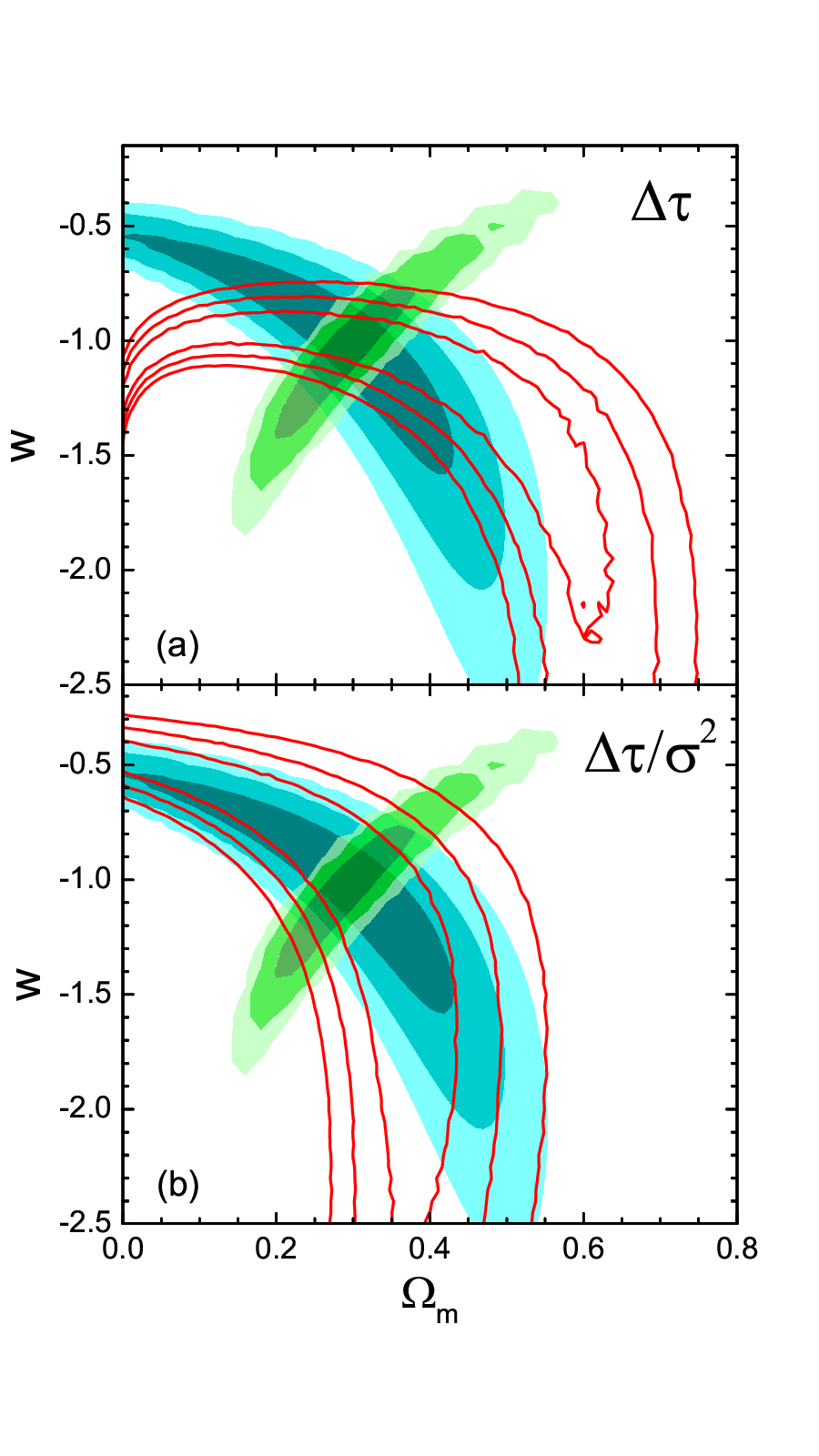}
\vskip-0.4in
\caption{Constraint results for the $w$CDM model using 50 lensed GW-EM systems (red solid lines),
compared with those associated with the 580 Union2.1 SNe Ia data (cyan contours) and CMB data (green contours):
(a) simulations for the $\Delta\tau$ method; (b) simulations for the $\Delta\tau/\sigma^{2}$ method.}
\label{fig:f4}
\vskip-0.3in
\end{center}
\end{figure}

\begin{figure}
\begin{center}
\vskip-0.3in
\includegraphics[width=0.45\textwidth]{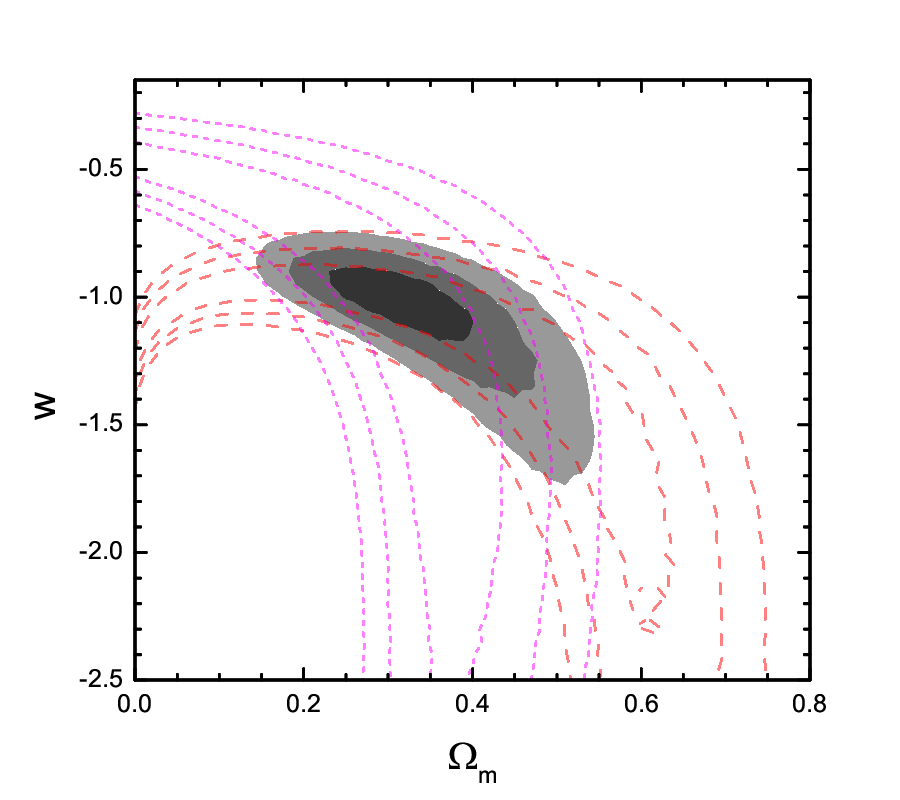}
\vskip-0.1in
\caption{Cosmological constraints on the $w$CDM model from 50 lensed GW-EM systems
for three different cases: $\Delta\tau$ (red dashed lines), $\Delta\tau/\sigma^{2}$ (magenta dot lines)
and $\Delta\tau + \Delta\tau/\sigma^{2}$ (black contours).}
\label{fig:f5}
\vskip-0.3in
\end{center}
\end{figure}

For the $w$CDM model, $w$ is constant but possibly different from $-1$. For a flat Universe ($\Omega_{k}=0$),
there are three free parameters: $\Omega_{\rm m}$, $w$, and $H_{0}$. Here, we marginalize $H_{0}$ in the $w$CDM model
to find the confidence levels in the $\Omega_{\rm m}-w$ plane. We demonstrate that lensed GW-EM systems can be a viable way
to constrain the dark-energy equation of state. To gauge the impact of these constraints more clearly, we show in Fig.~\ref{fig:f4}
the confidence regions (red solid lines) for $\Omega_{\rm m}$ and $w$ using 50 simulated strongly lensed GW-EM systems,
and we compare these to the constraint contours for the 580 Union2.1 SNe Ia data \citep{2012ApJ...746...85S} (represented by the cyan
contours in Fig.~\ref{fig:f4}). It is straightforward to see how effectively the lensed GW-EM systems could be used as a cosmological probe.
With a sample size of $\sim50$, the contour size of lensed GW-EM systems is already comparable to that of 580 SNe Ia data.
Furthermore, we note that the constraints obtained from the $\Delta\tau$ method (Fig.~\ref{fig:f4}(a)) and the $\Delta\tau/\sigma^{2}$ method
(Fig.~\ref{fig:f4}(b)) are intersecting, much better constraints can be achieved when combining these two methods (see the black contours in Fig.~\ref{fig:f5}).
That is, if we are lucky enough to have the joint measurements of $\Delta\tau$ and $\sigma^{2}$ for each lens, the better cosmological constraints
would be obtained by combining these two methods.

To illustrate the degeneracy breaking power of the proposed methods, we also plot the constraint contours of CMB from \emph{Planck} 2015 measurements
(\citealt{2016A&A...594A..14P}, see \citealt{2017arXiv170803143W} for more details on the calculations of CMB) (represented by the green contours
in Figs.~\ref{fig:f3} and \ref{fig:f4})\footnote{For the $w$CDM model, because CMB observables explicitly depend on $H_{0}$, we use a Gaussian prior
on its value, $H_{0}=67.8\pm0.9$ km $\rm s^{-1}$ $\rm Mpc^{-1}$, to guide the minimization procedure over $H_{0}$.}.
We can see that when the observations of lensed GW-EM systems are verified in the future, CMB constraints
would benefit from having the constraints of lensed GW-EM systems overlaid.

\section{Summary and discussion}
\label{sec:summary}

Although the constraints on cosmological parameters have reached a high precision, all of the constraints so far have relied on
EM observations alone. New multimessenger signals exploiting different emission channels are worth exploring in cosmology.
Recently, \cite{2017arXiv170304151L} proposed that future time-delay measurements of strongly lensed GW signals
and their EM counterparts have great potential to infer cosmological parameters. Compared with the traditional approach of using
strongly lensed quasar systems observed in the EM domain, the approach with lensed systems observed in both GW and EM windows
has two advantages in constraining cosmological parameters. First, the time delays ($\Delta\tau$) between lensed images
inferred from the GW signals would reach an extremely high accuracy ($\sim0.1$ s) from the detection pipeline, and such accurate
measurements of $\Delta\tau$ would play an important role in boosting the development of precision cosmology. Secondly,
with gravitationally lensed GWs and EM signals, the accuracy of lens modelling could be improved to some extent,
leading to better constraints on cosmological parameters.

In the EM window, \cite{2009A&A...507L..49P} suggested that the joint observations of the time delay ($\Delta\tau$) between
lensed quasar images and the velocity dispersion ($\sigma$) of the lensing galaxy are more effective to constrain cosmological parameters
than $\Delta\tau$ or $\sigma^{2}$ separately. In this work, we apply the $\Delta\tau/\sigma^{2}$ method, for the first time, to the strongly lensed
systems observed in both GW and EM windows. We prove that both $\Delta\tau$ and $\Delta\tau/\sigma^{2}$ from strongly lensed
GW-EM systems can serve as powerful cosmic rulers. From the comparison of the two different methods, we confirm that
the $\Delta\tau/\sigma^{2}$ method can provide tighter constraints on $\Omega_{\rm m}$ than the $\Delta\tau$ method,
(i.e. using $\Delta\tau/\sigma^{2}$ can make it easier to differentiate different cosmological models).
Furthermore, we show that with a moderate sample size of $\sim 50$, a constraint on the dark energy equation of state $w$
can be reached that is comparable to the 580 Union2.1 SNe Ia sample. Combining the $\Delta\tau$ and $\Delta\tau/\sigma^{2}$ methods,
it is possible to achieve higher accuracy in constraining $w$.

The recent Advanced LIGO observations of binary black hole mergers GW150914 \citep{2016PhRvL.116f1102A}, GW151226 \citep{2016PhRvL.116x1103A},
and GW170104 \citep{2017PhRvL.118v1101A} have initiated the era of GW astronomy. Because of the high sensitivity,
the planned third-generation GW detectors, such as the ET, could observe strongly lensed GWs.
Recent works \citep{2013JCAP...10..022P,2014JCAP...10..080B,2015JCAP...12..006D} have carefully studied the prospects of observing
strongly lensed GWs from merging double compact objects, which predicted that the ET would detect about 50--100 strongly lensed GW events per year.
Although a considerable catalogue of lensed GWs would be obtained, the measurements of strongly lensed GW-EM systems suggested by our method
will still be extremely hard in practice. The measurements must meet three requirements: (i) we need an EM counterpart to give the exact location
of the lensed images; (ii) we need to get a source redshift from that EM counterpart; (iii) we need the GW source to have a detectable host galaxy
so we can carry out detailed lens modelling. It is not clear what fraction of lensed GW events will actually satisfy these three requirements.
If, in the future, gravitationally lensed GWs and their EM counterparts are detected simultaneously,
the prospects for the study of cosmology with such lensing systems, as discussed in this work, will be very promising.

\section*{Acknowledgements}
We are grateful to the anonymous referee for insightful comments that have
helped us improve the presentation of the paper. We also thank Kai Liao for his kind assistance.
This work is partially supported by the National Basic Research Program (``973'' Program)
of China (Grant No. 2014CB845800), the National Natural Science Foundation of China
(Grant Nos. 11673068, 11603076, 11433009, and 11373068), the Youth Innovation Promotion
Association (2011231 and 2017366), the Key Research Program of Frontier Sciences (QYZDB-SSW-SYS005),
the Strategic Priority Research Program ``Multi-waveband gravitational wave Universe''
(Grant No. XDB23000000) of the Chinese Academy of Sciences, the Natural Science Foundation
of Jiangsu Province (Grant No. BK20161096), and the Guangxi Key Laboratory for Relativistic Astrophysics.







\bsp	
\label{lastpage}
\end{document}